\documentstyle[12pt]{article}
\newcommand{\beq}{\begin{equation}}
\newcommand{\eeq}{\end{equation}}
\newcommand{\beqa}{\begin{eqnarray}}
\newcommand{\eeqa}{\end{eqnarray}}
\newcommand{\ba}{\begin{array}}
\newcommand{\ea}{\end{array}}
\newcommand{\CR}{\nonumber \\}
\newcommand{\pa}{\partial}

\newcommand{\La}{\Lambda}

\newcommand{\bra}{\langle}
\newcommand{\ket}{\rangle}

\newcommand{\rep}{{\cal R}}
\newcommand{\half}{{1\over 2}}

\begin{document}

\begin{titlepage}
\null
\begin{flushright} 
hep-th/9708017  \\
UTHEP-367  \\
August, 1997
\end{flushright}
\vspace{0.5cm} 
\begin{center}
{\Large \bf
Flat Coordinates, Topological Landau-Ginzburg Models and 
the Seiberg-Witten Period Integrals
\par}
\lineskip .75em
\vskip2.5cm
\normalsize
{\large Katsushi Ito and Sung-Kil Yang} 
\vskip 1.5em
{\large \it Institute of Physics, University of Tsukuba, Ibaraki 305, Japan}
\vskip3cm
{\bf Abstract}
\end{center} \par
We study the Picard-Fuchs differential equations for the Seiberg-Witten period
integrals in $N=2$ supersymmetric Yang-Mills theory. For A-D-E gauge groups 
we derive the Picard-Fuchs equations by using the flat coordinates in the 
A-D-E singularity theory. We then find that these are equivalent to the 
Gauss-Manin system for two-dimensional A-D-E topological Landau-Ginzburg 
models and the scaling relation for the Seiberg-Witten differential. 
This suggests an interesting relationship between four-dimensional $N=2$ 
gauge theories in the Coulomb branch and two-dimensional topological field 
theories.
\end{titlepage}

\baselineskip=0.7cm

The low-energy properties of the Coulomb branch of $N=2$ supersymmetric
gauge theories in four dimensions are exactly described by holomorphic data
associated with certain algebraic curves \cite{SW},\cite{review}. 
In particular, the BPS mass formula
is expressed in terms of the period integrals of the so-called
Seiberg-Witten (SW) differential $\lambda_{SW}$
\beq
a^i=\oint_{A_i}\lambda_{SW}, \hskip10mm {a_D}_i=\oint_{B_i}\lambda_{SW},
\hskip10mm i=1,\cdots ,r
\label{period}
\eeq
along one-cycles $A_i$, $B_i$ with appropriate intersections on the curve.
Here $r$ is the rank of the gauge group $G$.

In order to evaluate the period integrals (\ref{period}) an efficient way is
to employ the Picard-Fuchs equations obeyed by the periods
\cite{CeDaFe}-\cite{It}. 
The purpose of
this paper is to show that, for $N=2$ Yang-Mills theories with A-D-E gauge 
groups, these Picard-Fuchs equations are direct consequences of the
Gauss-Manin system for two-dimensional A-D-E topological Landau-Ginzburg 
models. 

A general scheme for SW curves in $N=2$ A-D-E Yang-Mills theories is proposed
in \cite{MW} in view of integrable systems \cite{Gor},\cite{NT},\cite{IM}.
According to \cite{MW}, when describing SW curves we first introduce
\beq
P_{\cal R}(x,u_i)={\rm det} (x-\Phi_{\cal R}).
\label{charapol}
\eeq
This is the characteristic polynomial in $x$ of order ${\rm dim}\, {\cal R}$
where ${\cal R}$ is an irreducible representation of $G$. 
Here $\Phi_{\cal R}$ is a representation
matrix of ${\cal R}$ and $u_i$ $(i=1,\cdots ,r)$ is the Casimir built out 
of $\Phi_{\cal R}$ whose degree equals $e_i+1$ with $e_i$ being the $i$-th
exponent of $G$. Especially $u_1$ denotes the quadratic Casimir and $u_r$
the top Casimir of degree $h$ where $h$ is the dual Coxeter number of $G$; 
$h=r+1, 2r-2, 12, 18,30$ for $G=A_r, D_r,
E_6, E_7, E_8$, respectively. The quantum SW curve is then given by
\beq
\widetilde P_{\cal R}(x,z,u_i) \equiv
P_{\cal R} \left( x,u_i+\delta_{i r} 
\left( z+\frac{\mu^2}{z}\right) \right) =0,
\label{curve}
\eeq
where $\mu^2=\La^{2 h}$ with $\Lambda$ being the dynamical scale and $u_i$
are now considered as gauge invariant moduli parameters in the Coulomb branch. 

The surface determined by (\ref{curve}) is viewed as a multi-sheeted foliation
$x(z)$ over the base ${\bf CP}^1$ with coordinate $z$. For (\ref{curve})
the SW differential takes the form
\beq
\lambda_{SW}=x\, {dz \over z}
\label{swd}
\eeq
For a higher-dimensional representation the surface becomes quite complicated.
However, the physics of $N=2$ Yang-Mills theory is described by a complex
rank $G$-dimensional sub-variety of the Jacobian. It is argued in \cite{MW}
that this sub-variety is the special Prym variety which is universal and
independent of the representation ${\cal R}$. We thus expect that the
Picard-Fuchs equations for a given curve should take the same form irrespective
of the representation $\rep$ as a result of the universality of the
special Prym. This is indeed the case as we will see momentarily.

Before discussing the Picard-Fuchs equations we need to see how the SW curves
are related to the topological Landau-Ginzburg (LG) models in two dimensions.
We first express the curve (\ref{curve}) in the form
\beq
z+{\mu^2 \over z}+u_r=\widetilde W_G^\rep (x,u_1,\cdots ,u_{r-1}).
\label{solve}
\eeq
When a representation becomes higher we cannot write down 
$\widetilde W_G^\rep$ in a closed
form, but it is always possible to find $\widetilde W_G^\rep$ as a series
expansion around $x=\infty$. Notice that the overall degree of 
$\widetilde W_G^\rep$ is equal to $h$.

For the fundamental representation of $A_r$ and $D_r$ we have
\beqa
&& \widetilde W_{A_r}^{\bf r+1}=x^{r+1}-u_1x^r- \cdots -u_{r-1}x, \CR
&& \widetilde W_{D_r}^{\bf 2r}=x^{2r-2}-u_1x^{2r-4}- \cdots -u_{r-2}x^2
-{u_{r-1}^2 \over x^2},
\eeqa
where $u_{r-1}$ in $D_r$ theory is the degree $r$ Casimir (Pfaffian invariant).
If we define
\beq
W_G^\rep (x,u_1,\cdots, u_r)=\widetilde W_G^\rep (x,u_1,\cdots, u_{r-1})-u_r,
\label{sp}
\eeq
then $W_{A_r}^{\bf r+1}$ and $W_{D_r}^{\bf 2r}$ are the familiar LG
superpotentials for the $A_r$- and $D_r$-type topological minimal models,
respectively \cite{DVV}. 

For the fundamental representation ${\bf 27}$ of $E_6$ the explicit form of
(\ref{sp}) is obtained in \cite{LW}
\beq
W_{E_6}^{\bf 27}={1\over x^3} \left( q_1 \pm p_1\sqrt{p_2}\right)-u_6 ,
\label{e6curve}
\eeq
where
\beqa
&&q_1=270 x^{15}+342 u_{1} x^{13}+162 u_{1}^2 x^{11}-252 u_{2} x^{10}
+(26 u_{1}^3+18 u_{3}) x^{9}-162 u_{1} u_{2} x^{8} \CR
&& \quad \quad  +(6 u_{1} u_{3} -27 u_{4}) x^{7}
-(30 u_{1}^2 u_{2}-36 u_{5}) x^{6}
+(27 u_{2}^2 -9 u_{1} u_{4}) x^{5} \CR
&& \quad \quad -(3 u_{2} u_{3}-6 u_{1} u_{5}) x^{4}
-3 u_{1} u_{2}^2 x^3-3 u_{2} u_{5} x-u_{2}^3, \CR
&& p_1=78 x^{10}+60 u_{1} x^{8} +14 u_{1}^2 x^{6}-33 u_{2} x^{5}
+2 u_{3} x^{4}-5 u_{1} u_{2} x^{3}-u_{4} x^{2}-u_{5} x-u_{2}^2,  \CR
&& p_2=12 x^{10}+12 u_{1} x^{8}+4 u_{1}^2 x^{6}-12 u_{2} x^{5}+u_{3}x^{4}
-4 u_{1} u_{2} x^{3}-2 u_{4} x^{2}+4 u_{5} x+u_{2}^2
\eeqa
and the degrees of Casimirs are $\{2,5,6,8,9,12 \}$.
In the standard description the LG superpotential for the topological $E_6$
minimal model is given by versal deformations of the $E_6$ singularity
written in terms of two variables (plus one gaussian variable) \cite{MVW}. 
It is, however, found in
\cite{EY} that $W_{E_6}^{\bf 27}$ is a single-variable version of the
LG superpotential for the $E_6$ model and gives a new description of the
$E_6$ singularity.

The other interesting example studied in \cite{EY} is the curve in the
ten-dimensional representation ${\bf 10}$ for $A_4$ Yang-Mills theory.
One has \cite{MW}
\beq
W_{A_4}^{\bf 10}=\half \left( q_1 \pm p_1\sqrt{p_2} \right)-u_4,
\eeq
where
\beqa
&&q_1=11 x^5+4 u_1 x^3+7 u_2 x^2+(u_1^2-4 u_3) x+u_1 u_2, \CR
&&p_1=5 x^3+u_1 x+u_2,\CR
&&p_2=5 x^4+2 u_1 x^2+4 u_2 x+u_1^2-4 u_3.
\eeqa
It is then shown that the square-root-type superpotential $W_{A_4}^{\bf 10}$
reproduces the correct result for the topological $A_4$ LG model which
is usually described in terms of the polynomial-type superpotential
$W_{A_4}^{\bf 5}$.

On the basis of the above observations we assume in the following that the
function $W_G^\rep$ in (\ref{sp}) for $G=A, D, E$ is identified as the
LG superpotential for the topological A-D-E minimal models irrespective
of the representation $\rep$. The validity of this assumption may be
ensured by the universality of the special Prym.

In two-dimensional topological field theories the flat coordinates for the
moduli space play a distinguished role 
\cite{Sa},\cite{No},\cite{DVV},\cite{BV}. For a LG model with the 
superpotential (\ref{sp}), following \cite{EYY}, we fix the flat coordinate 
$t_i$ by the residue integral
\beq
t_i=c_i \oint dx W_G^\rep(x,u)^{e_i \over h}, \hskip10mm i=1,\cdots, r
\label{flat}
\eeq
with a normalization constant $c_i$ which will be specified later. 
Here $e_i$ denotes the exponents of $G$ and the
residue integral is taken around $x=\infty$. \footnote[2]{In $D_r$ theory we 
have to consider the residue integral around $x=0$ as well. See \cite{EYY}.}
The flat coordinates obtained from (\ref{flat}) are polynomials in $u_j$.
For instance, the $E_6$ superpotential (\ref{e6curve}) with the $+$ sign
chosen yields \cite{EY} \footnote[3]{Our normalization of $u_6$ here differs
by a factor of 3456 from \cite{EY}.}
\beqa
&& t_1= -{3^{2/3}\over 12 } u_{1}, \CR
&& t_2={\sqrt{2}\, 3^{2/3}\over 24} u_{2}, \CR
&& t_3=-{1\over 192} \left(u_{3}-{1\over2} u_{1}^3\right), \CR
&& t_4=-
{3^{2/3}\over 288}\left( u_{4}+{1\over 8} u_{1} u_{3}
-{5\over 96} u_{1}^4\right),
\CR
&& t_5={\sqrt{2}\over 96} u_{5}, \CR
&& t_6=-u_6-{1\over 3456}\left(
 {3\over 64}u_{3}^2+{9\over2}u_{1} u_{2}^2
-{3\over8} u_{1}^2 u_{4}-{5\over 64} u_{1}^3 u_{3}+{3\over 127} u_{1}^6
\right).
\label{e6flat}
\eeqa

We define the primary fields $\phi_i^\rep$ as the derivatives of the
superpotential
\beq
\phi_i^\rep (x)={\partial W_G^\rep(x,u) \over \partial t_i}, 
\hskip10mm i=1,\cdots, r
\eeq
among which the field $\phi_r^\rep =1$ is the identity operator, {\it i.e.}
the puncture operator $P$ in the context of two-dimensional topological
gravity. The one-point functions of the $n$-th gravitational descendant
$\sigma_n (\phi_i^\rep)$ are evaluated by the residue integrals \cite{EYY}
\beq
\bra \sigma_n (\phi_i^\rep) \ket 
= b_{n,i}\sum_{j=1}^r \eta_{ij} \oint W_G^\rep(x,u)^{{e_j \over h} +n+1},
\hskip10mm n=0,1,2, \cdots
\eeq
with $b_{n,i}$ being certain constant. Here the topological metric $\eta_{ij}$
is given by
\beqa
\eta_{ij}&=& \bra \phi_i^\rep \phi_j^\rep P \ket \CR
&=& b_{0,r} {\partial^2 \over \partial t_i \partial t_j}
\oint W_G^\rep(x,u)^{1+1/h}.
\eeqa
Note that independence of $\eta_{ij}$ on the moduli parameters is 
characteristic of the flat coordinate system. The charge selection rule 
indicates $\eta_{ij} \propto \delta_{e_i+e_j, h}$. We adjust normalization
constants $c_i, \, b_{n,i}$ in such a way that 
$\eta_{ij} =\delta_{e_i+e_j, h}$.

The important property of the primary fields is that they generate the
closed operator algebra
\beq
\phi_i^\rep (x)\, \phi_j^\rep (x)=\sum_{k=1}^r {C_{ij}}^k(t)\, \phi_k^\rep (x)
+Q_{ij}^\rep(x) \, \partial_x  W_G^\rep(x),
\label{OPE}
\eeq
where $Q_{ij}^\rep$ are obtained from
\beq
{\partial^2 W_G^\rep(x) \over \partial t_i \partial t_j}
=\partial_x \, Q_{ij}^\rep(x).
\label{contact}
\eeq
The relation (\ref{contact}) is well-known for superpotentials
$W_{A_r}^{\bf r+1}$ and $W_{D_r}^{\bf 2r}$ as the key property of the flat
coordinates in the singularity theory. What is remarkable is that 
(\ref{contact}) also holds for superpotentials $W_{E_6}^{\bf 27}$ and
$W_{A_4}^{\bf 10}$ of square-root type \cite{EY}. Hence we assume that 
(\ref{contact}) is valid for any representation $\rep$. 

According to our universality assumption the structure constants ${C_{ij}}^k$
are independent of $\rep$ since $C_{ijk}={C_{ij}}^\ell \eta_{\ell k}$ are
given by the three-point functions
\beq
C_{ijk}(t)=\bra \phi_i^\rep \phi_j^\rep \phi_k^\rep \ket
\eeq
which are topological-invariant physical observables. In two-dimensional
topological theories all of the topological correlation functions are
completely determined by the free energy $F(t)$. The three-point functions
are then given by 
$C_{ijk}(t)= \pa^3 F(t)/ \pa t_i \pa t_j \pa t_k$.
For the $E_6$ model the free energy is known to be \cite{KlThSc}
\beqa
&&F_{E_6}(t)=t_{6}t_{4}t_{3}+t_{6}t_{5}t_{2}
+{t_{6}^2t_{1} \over 2}+{t_{5}^2T_{4} \over 2}
+{t_{5}t_{3}^2t_2 \over 2}+{t_{4}^2t_2^2 \over 4} \CR
&& \hskip15mm 
+{t_{3}t_{2}^4 \over 12}+{t_{4}^3t_{1} \over 6}+{t_{5}^2t_{3}t_{1} \over 2}
+{t_{3}^4t_{1} \over 12}+{t_{4}t_{3}t_{2}^2t_{1} \over 2}
+{t_{5}t_{2}^3t_{1} \over 6} \CR
&& \hskip15mm
+{t_{5}t_{4}t_{2}t_{1}^2 \over 2}+{t_{3}^2t_{2}^2t_{1}^2 \over 4}
+{t_{4}t_{3}^2t_{1}^3 \over 6}+{t_{5}t_{3}t_{2}t_{1}^3 \over 6}
+{t_{2}^4t_{1}^3 \over 24}+{t_{5}^2t_{1}^4 \over 24} \CR
&& \hskip15mm
+{t_{4}t_{2}^2t_{1}^4 \over 24}+{t_{4}^2t_{1}^5 \over 60}
+{t_{3} t_{2}^2t_{1}^5 \over 24}+{t_{3}^2t_{1}^7 \over 252}
+{t_{2}^2t_{1}^8 \over 3^22^6}+{t_{1}^{13} \over 11\cdot 13\cdot2^43^4}.
\label{e6free}
\eeqa

We are now ready to derive the Picard-Fuchs equations in $N=2$ A-D-E Yang-Mills
theories. It is easily seen from (\ref{swd}) and (\ref{solve}) that the SW
differential becomes
\beq
\lambda_{SW}={x\partial_{x} W \over \sqrt{W^2-4\mu^2}}\, dx,
\eeq
where the superpotential $W_G^\rep$ is denoted as $W$ for notational
simplicity. Similarly we obtain
\beq
{\partial \lambda_{SW} \over \partial t_i}
=-{1 \over \sqrt{W^2-4\mu^2}}\, {\partial W\over \partial t_i}\, dx
+d \left({x\over \sqrt{W^2-4\mu^2}}\, {\partial W\over \partial t_i} \right).
\label{derivi}
\eeq
Total derivative terms will be suppressed in what follows.

Let us first note that the superpotential $W$ possesses the quasihomogeneous
property, leading to the scaling relation
\beq
x\partial_x W+\sum_{i=1}^r q_it_i{\partial W \over \partial t_i} =hW,
\eeq
where
we have used the flat coordinates $\{ t_i \}$ instead of $\{ u_i \}$ and
$q_i=e_i+1$ is the degree of $t_i$. Then one finds
\beq
\lambda_{SW}-\sum_{i=1}^{r} q_it_i{\partial \lambda_{SW} \over \partial t_i}=
{hW \over \sqrt{W^2-4\mu^2}}.
\eeq
Acting the Euler derivative $\sum_j q_jt_j {\partial \over \partial t_j}$
on both sides we get 
\beqa
&& \hskip5mm \sum_{i,j=1}^{r} q_iq_jt_it_j 
{\partial \lambda_{SW} \over \partial t_i \partial t_j}
+\sum_{i=1}^{r} q_i(q_i-2)t_i{\partial \lambda_{SW} \over \partial t_i} 
+\lambda_{SW}  \CR
&& = 4\mu^2h^2 {Wdx \over (W^2-4\mu^2)^{3/2}}
+{hWdx \over \sqrt{W^2-4\mu^2}}
-4\mu^2 h {x \partial_x Wdx \over (W^2-4\mu^2)^{3/2}}.
\eeqa
The rhs is expressed as
\beq
4\mu^2h^2{\partial^2 \lambda_{SW} \over \partial t_r^2}
+h d \left( {xW \over \sqrt{W^2-4\mu^2}} \right) ,
\eeq
and hence
\beq
\sum_{i,j=1}^{r} q_iq_jt_it_j
{\partial \lambda_{SW} \over \partial t_i \partial t_j}
-4\mu^2h^2{\partial^2 \lambda_{SW} \over \partial t_r^2}
+\sum_{i=1}^{r} q_i(q_i-2)t_i{\partial \lambda_{SW} \over \partial t_i}
+\lambda_{SW} =0.
\eeq
This is rewritten as
\beq
\left( \sum_{i=1}^{r} q_it_i{\partial \over \partial t_i}
-1 \right)^2 \lambda_{SW}
-4\mu^2h^2{\partial^2 \lambda_{SW} \over \partial t_r^2}=0.
\label{PFla1}
\eeq
The second term represents the scaling violation due to the dynamically 
generated scale $\mu^2=\Lambda^{2h}$ since (\ref{PFla1}) reduces to the
scaling relation for $\lambda_{SW}$ in the classical limit 
$\mu^2 \rightarrow 0$. Note that if we think of $\lambda_{SW}$ as a function
of $t_i$ and $\mu$ the fact that $\lambda_{SW}$ is of degree ($=$ mass
dimension) one implies 
\beq
\left( \sum_{i=1}^{r} q_it_i{\partial \over \partial t_i}
+h \mu {\pa \over \pa \mu} -1 \right) \lambda_{SW}=0
\eeq
{}from which one readily recovers (\ref{PFla1}).

The other set of differential equations for $\lambda_{SW}$ is obtained with
the aid of the operator product expansion (\ref{OPE}). Using 
(\ref{OPE}), (\ref{contact}) and (\ref{derivi}) it is immediate
to follow
\beqa
{\pa^{2}\over \pa t_{i}\pa t_{j}}\lambda_{SW}
&=& -{1\over \sqrt{W^2-4\mu^2}} {\pa^{2} W\over \pa t_{i}\pa t_{j}}\, dx
+{\phi_{i}\phi_{j} W \over (W^2-4\mu^2)^{3/2}}\, dx  \CR
&=& -{1\over \sqrt{W^2-4\mu^2}}\, d Q_{ij}
+{ (\sum_{k} {C_{i j}}^{k} \phi_{k}+Q_{i j} \pa_{x}W) W
\over (W^2-4\mu^2)^{3/2}}\, dx \CR
&=&\sum_{k} { {C_{i j}}^{k} W
\over (W^2-4\mu^2)^{3/2}} {\pa W\over \pa t_{k}}\, dx
-d \left( {Q_{i j}\over \sqrt{W^2-4\mu^2}}\right).
\eeqa
Notice that
\beq
{ W \over (W^2-4\mu^2)^{3/2}} {\pa W\over \pa t_{k}}\, dx
={\pa^{2}\over \pa t_{k}\pa t_{r}} \lambda_{SW},
\eeq
thereby we find
\beq
{\pa^{2}\over \pa t_{i}\pa t_{j}}\lambda_{SW}
=\sum_{k} {C_{i j}}^{k}(t) {\pa^{2}\over \pa t_{k}\pa t_{r}} \lambda_{SW}.
\eeq

The Picard-Fuchs equations for the SW period integrals 
$\Pi =\oint \lambda_{SW}$ now read
\beqa
\left( \sum_{i=1}^{r}q_{i}t_{i}{\pa\over \pa t_{i}}-1 \right)^2\Pi
-4\mu^2 h^{2}{\pa^{2}\Pi\over \pa t_{r}^2}=0, \CR
{\pa^{2}\Pi\over \pa t_{i}\pa t_{j}}
-\sum_{k=1}^r {C_{i j}}^{k}(t) {\pa^{2}\Pi \over \pa t_{k}\pa t_{r}}=0.
\label{PFflat}
\eeqa
We observe here that
the second equations are nothing but the Gauss-Manin differential equations 
for period integrals expressed in the flat coordinate in topological LG 
models \cite{EYY}. Converting the flat coordinate $\{ t_{i} \}$ into the 
standard moduli parameters $\{ u_{i} \}$ is straightforward. 
One obtains
\beqa
{\cal L}_{0}\Pi&\equiv&
\left( \sum_{i=1}^{r}q_{i}u_{i}{\pa\over \pa u_{i}}-1 \right)^2\Pi
-4\mu^2 h^{2}{\pa^{2}\Pi\over \pa u_{r}^2}=0, \CR
{\cal L}_{i j}\Pi&\equiv&{\pa^{2}\Pi\over \pa u_{i} \pa u_{j}}
+\sum_{k=1}^{r} A_{i j k} (u)\, {\pa^{2}\Pi\over \pa u_{k} \pa u_{r}}
+\sum_{k=1}^{r}B_{i j k}(u)\, {\pa \Pi\over \pa u_{k}}=0,
\label{PFfinal}
\eeqa
where $\pa u_k/\pa t_r=-\delta_{kr}$ and 
the coefficient functions are given by
\beqa
A_{ijk}(u)&=& \sum_{m,n,\ell =1}^r {\pa t_{m}\over \pa u_{i}}
  {\pa t_{n}\over\pa u_{j}}
 {\pa u_{k}\over \pa t_{\ell}}\, {C_{m n}}^{\ell}(u), \CR
B_{i j k}(u)&=& -\sum_{n=1}^{r} {\pa^{2}t_{n}\over \pa u_{i}\pa u_{j}}
{\pa u_{k}\over \pa t_{n}},
\eeqa
which are all polynomials in $u_i$.
It is interesting that the Picard-Fuchs equations in four-dimensional 
$N=2$ Yang-Mills theory are essentially governed by the data
in two-dimensional topological LG models.

Let us present some examples of the Picard-Fuchs equations.
For $A_{r}$ $(r\leq 4$) case, applying the flat coordinates for topological
$A_{r}$ LG  models \cite{DVV}, one can reproduce from (\ref{PFfinal}) 
the Picard-Fuchs equations of $A_{r}$-type gauge groups derived earlier
by several authors \cite{CeDaFe},\cite{KlLeTh},\cite{ItSa},\cite{Al}.
As a more nontrivial example we write down explicitly the Picard-Fuchs 
equations for $N=2$ $E_{6}$ Yang-Mills theory. 
Using (\ref{e6flat}) and (\ref{e6free}) we obtain from (\ref{PFfinal}) that
\beqa
{\cal L}_{0}&=&
4 u_{1}^{2}\pa_{1}^2
+20 u_{1} u_{2} \pa_{1}\pa_{2}
+24 u_{1} u_{3} \pa_{1}\pa_{3}
+32 u_{1} u_{4} \pa_{1}\pa_{4}
+36 u_{1} u_{5} \pa_{1}\pa_{5}
+48 u_{1} u_{6} \pa_{1}\pa_{6} \CR
& & +25 u_{2}^2 \pa_{2}^2
+60 u_{2} u_{3} \pa_{2}\pa_{3}
+80 u_{2} u_{4} \pa_{2}\pa_{4}
+90 u_{2} u_{5} \pa_{2}\pa_{5}
+120 u_{2} u_{6} \pa_{2}\pa_{6} \CR
& & +36 u_{3}^2\pa_{3}^2
+96 u_{3} u_{4} \pa_{3}\pa_{4}
+108 u_{3} u_{5}\pa_{3}\pa_{5}
+144 u_{3} u_{6} \pa_{3} \pa_{6} \CR
& & +64 u_{4}^2 \pa_{4}^2
+144 u_{4} u_{5}\pa_{4}\pa_{5}
+192 u_{4} u_{6}\pa_{4}\pa_{6}
+162 u_{5}^2 \pa_{5}^2
+216 u_{5} u_{6}\pa_{5}\pa_{6} \CR
& & +144(u_{6}^2-4 \mu^2)\pa_{6}^2 
+15 u_{2}\pa_{2}
+24  u_{3}\pa_{3}
+48  u_{4}\pa_{4}
+63  u_{5}\pa_{5}
+120  u_{6}\pa_{6}+1,  \CR
\widehat{\cal L}_{5 5}&=&\pa_{5}^2-144 \pa_{3}\pa_{6}+36u_{1}\pa_{4}\pa_{6}
+27u_{3}\pa_{6}^2,
\CR
\widehat{\cal L}_{2 6}&=& \pa_{2}\pa_{6}-{1\over3}\pa_{4}\pa_{5},
\CR
\widehat{\cal L}_{4 4 }&=&\pa_{4}^2+6u_{1}\pa_{3}\pa_{6}+{3\over2}u_{2}\pa_{5}\pa_{6}
+{9\over2}u_{4}\pa_{6}^2,
\CR
\widehat{\cal L}_{3 5}&=& \pa_{3}\pa_{5}+{9\over2}u_{2}\pa_{4}\pa_{6}
+{27\over 2}u_{5}\pa_{6}^2,
\CR
\widehat{\cal L}_{3  4}&=&\pa_{3}\pa_{4}-{3\over4}\pa_{1}\pa_{6},
\CR
\widehat{\cal L}_{1 6}&=&\pa_{1}\pa_{6}-{1\over36}\pa_{2}\pa_{5}
-{1\over3}u_{1}\pa_{4}^2-{1\over4}u_{3}\pa_{4}\pa_{6},
\CR
\widehat{\cal L}_{2 4}&=&\pa_{2}\pa_{4}+2u_{1}\pa_{3}\pa_{5}+{1\over2}u_{2}\pa_{5}^2
+{3\over2}u_{4}\pa_{5}\pa_{6},
\CR
\widehat{\cal L}_{3 3 }&=& \pa_{3}^2-{3\over16}u_{1}\pa_{1}\pa_{6}
+{3\over 32} u_{2}\pa_{2}\pa_{6}-{3\over 16}u_{3}\pa_{3}\pa_{6}+{3\over 32}
u_{5}\pa_{5}\pa_{6}-{3\over 32}\pa_{6}, \CR
\widehat{\cal L}_{2 3}&=& \pa_{2}\pa_{3}-{1\over 4}\pa_{1}\pa_{5}, \CR
\widehat{\cal L}_{1 5 }&=& \pa_{1}\pa_{5}+6u_{2}\pa_{4}^2+18 u_{5}\pa_{4}\pa_{6},
\CR
\widehat{\cal L}_{2 2}&=& \pa_{2}^2-12\pa_{1}\pa_{4}-18 u_{1}^2\pa_{1}\pa_{6}
-18 u_{1}u_{2}\pa_{2}\pa_{6}+3 u_{3}\pa_{4}^2-18 u_{1}u_{4}\pa_{4}\pa_{6}
-18 u_{1}\pa_{6},
\CR
\widehat{\cal L}_{1 4}&=&\pa_{1}\pa_{4}+8 u_{1}\pa_{3}^2+2 u_{2}\pa_{3}\pa_{5}
+6 u_{4}\pa_{3}\pa_{6},
\CR
\widehat{\cal L}_{1 3}&=&\pa_{1}\pa_{3}-{1\over4}u_{1}\pa_{1}\pa_{4}
+{1\over8}u_{2}\pa_{2}\pa_{4}-{1\over4}u_{3}\pa_{3}\pa_{4}
+{1\over 8}u_{5}\pa_{4}\pa_{5}-{1\over8}\pa_{4},
\CR
\widehat{\cal L}_{1 2 }&=&\pa_{1}\pa_{2}-9 u_{1}u_{2}\pa_{1}\pa_{6}
-9 u_{2}^2\pa_{2}\pa_{6}-36 u_{1}u_{5}\pa_{3}\pa_{6}
-9 u_{2} u_{5}\pa_{5}\pa_{6}+2 u_{4}\pa_{3}\pa_{5}-9 u_{2}\pa_{6},
\CR
\widehat{\cal L}_{1 1}&=&\pa_{1}^2
+2 u_{1}^2\pa_{1}\pa_{3}
-{1\over4}u_{3}\pa_{1}\pa_{4}
+{1\over 4} u_{1}u_{2}\pa_{1}\pa_{5}
+2 u_{1}u_{4}\pa_{3}\pa_{4}
+{1\over2} u_{5}\pa_{2}\pa_{4}
-{1\over 4} u_{2}^2\pa_{2}\pa_{5} 
\CR
& & -{1\over 4} u_{2}u_{4}\pa_{4}\pa_{5} 
 +3 u_{1}\pa_{3}+{3\over4} u_{4}\pa_{6},
\label{eq:e6pf}
\eeqa
where $\pa_{i}={\pa\over \pa u_{i}}$ and 
$\widehat{\cal L}_{ij}={\cal L}_{i j}+(\mbox{linear combination of 
${\cal L}$'s})$.
We have also verified these equations directly from (\ref{swd})
and (\ref{e6curve}) without relying on the topological LG data. 
Computations are done by finding the relation
\beq
{\cal L} \lambda_{SW}
=d\left({f(x)+g(x)\sqrt{p_2}\over
\sqrt{p_2} \sqrt{x^{6}\left( ({W}_{E_{6}}^{\bf 27})^2
-4\mu^{2}\right) }} \right) 
\eeq
for a differential operator ${\cal L}$ in (\ref{eq:e6pf}).
Here $f(x)$ and $g(x)$ are polynomials in $x$ with order less than 22 and
17 respectively.

We have seen that the Picard-Fuchs equations in $N=2$ A-D-E Yang-Mills
theories are written in a concise universal form (\ref{PFflat}) in terms of
the flat coordinates and the topological LG data. They are equivalent to 
the scaling equation for the SW differential and the Gauss-Manin system for 
two-dimensional A-D-E topological LG models.
In the present work the flat coordinates for the moduli space and the
OPE relations (\ref{OPE}), (\ref{contact}) have played a crucial role.

There are several interesting issues deserved for further study.
First of all we may solve the Picard-Fuchs
equations (\ref{PFfinal}) in the weak-coupling region and compute
multi-instanton corrections to the prepotential for 
$E$-type gauge groups.
For $E$-type gauge groups, there is another proposal for the algebraic
curve \cite{AbAlGh} which has different strong-coupling physics from the
spectral curve.
By comparing the microscopic instanton results for these gauge groups
\cite{ItSa}, we may check which curves describe the correct 
non-perturbative structure of the low-energy effective theory.
In the case of exceptional gauge group $G_{2}$,
it is shown that the result from the spectral curve agrees with the 
microscopic instanton calculation \cite{It}. 
The analysis of the Gauss-Manin system for A-D-E
singularities would help in solving the problem \cite{No}.

Secondly it will be a straightforward exercise to see how the present
calculations are modified when we take non-simply laced gauge groups. It is
particularly interesting to check if the LG OPE relation holds for 
non-simply laced superpotentials. 

Finally we wish to clarify a possible deeper relation between 
four-dimensional $N=2$ Yang-Mills theory and two-dimensional topological
theory. We suspect, for instance, that the WDVV-like equation for the
SW prepotential considered in \cite{MMM} has its origin in the associativity
of the operator algebra (\ref{OPE}) in topological LG models.

\vskip10mm

We would like to thank T. Eguchi for stimulating discussions.
The work of S.K.Y. was partly supported by Grant-in-Aid for Scientific
Research on Priority Areas (Physics of CP Violation) from the Ministry
of Education, Science, and Culture of Japan.

\newpage

%%%%%%%  References

\end{document}